\documentstyle[12pt]{article}

\begin{document}
\hfill{DOE-ER40757-039}

\hfill{IASSNS-HEP 95/20}

\hfill{CPP-94-3}

\hfill{June, 1994}

\vspace{12pt}

\centerline{{\bf Quarks in the Skyrme-'t Hooft-Witten Model}}

\bigskip

\centerline{L.C. Biedenharn}
\centerline{{\it Center for Particle Physics, Department of Physics,
University of Texas at Austin,}}
\centerline{{\it Austin, TX 78712}}

 \bigskip

\centerline{L.P. Horwitz$^{(a)}$}
\centerline{{\it School of Natural Sciences, Institute for Advanced Study}}
\centerline{{\it Princeton, NJ 08540}}

\begin{abstract}
The three-flavor Skyrme-'t Hooft-Witten model is interpreted in terms of
a quark-like substructure, leading to a new model of explicitly confined
color-free ``quarks'' reminiscent of Gell-Mann's original pre-color quarks,
but with unexpected and significant differences.

\smallskip

\noindent PACS numbers: 12.38.Aw,
12.4.Aa, 21.60.Fw\end{abstract}

\bigskip

The standard theory for strong interaction physics is, by consensus, QCD,
but the inherent difficulties in applying this theory to long-range
low-energy phenomena (meson-nuclear physics, say) have so far been
insuperable.  In the mid-seventies 't Hooft\cite{1} demonstrated that the
gauged color group $SU(N_c)$, with $N_c \rightarrow \infty$, provided
an approximate approach to strong interactions.  In this limit, mesons
have masses that scale as $(N_c)^0$ and the meson resonances are
narrow with widths that scale as $(N_c)^{-1}$.   In contrast, baryons
(which contain $N_c$ quarks) have masses that scale as $(N_c)^1$ with
sizes and shapes that have an $N_c$ independent limit.  Witten\cite{2}
recognized that, for baryons, these are characteristics of a soliton,
reviving an earlier (topological) strong interaction model of
Skyrme\cite{3}.

	The Skyrme-'t Hooft-Witten (S'tW) model of baryons results in a classical
soliton solution of
the nonlinear chiral $SU(N_f) \times SU(N_f)$ model, from which the quantal
baryonic
states can be projected.  The crucial new ingredient in the S'tW model, due
to Witten,
is the {\it anomaly term}.  The implications of this anomaly term are
truly remarkable, and this term accounts for the impressive qualitative
(structural) agreements of the model with observation.  For two-flavors
the anomaly does not exist and the S'tW model is topologically trivial,
essentially equivalent, in fact, to the old strong-coupling spin-isospin
model.  Deeper insight into the structure of the two-flavor S'tW model
came from the quark hedgehog (large-$N_c$ quark) analysis\cite{4}.  This
analysis, combined with {\bf K}-symmetry, as discussed by Mattis and
Braaten\cite{5}, led to predictions for {\it two-flavor} meson-baryon
scattering in surprisingly good agreement with experiment.

The situation for the three-flavor S'tW model - - as a direct consequence
of the anomaly term - - is totally different.  Here the analysis in terms of
large-$N_c$ quarks is - - as we will prove - - incorrect in the literature.
Part of the problem is the understandable, but unfortunate, confusion
caused by denoting two distinct concepts by the same symbol.  The S'tW
model, when analyzed and interpreted in the way we propose, can be
seen as a model based on {\it quark-solitons}, which are generalized
(topological) ``quarks'' strongly reminiscent of Gell-Mann's original
(pre-color) ``mathematical'' quarks\cite{6} - - though not without some
unusual features of their own, as we shall show.

The three-flavor S'tW model is defined\cite{2} by the Poincar\'{e}
invariant action:
\begin{eqnarray}
S = \int d^4 xL + n \Gamma, (n \in Z),
\end{eqnarray}
where $\Gamma$ is the anomaly and $L$ is the Skyrme Lagrangian,
\begin{eqnarray}
L = \frac{F^2_\pi}{16} Tr \left\{ \left[ \partial_{\mu} U \partial^\mu
U \right] \right\} - \frac{1}{32e^2} Tr \left\{ \left[ ( \partial_{\mu} U)
U, (\partial_v U ) U \right]^2 \right\},
\end{eqnarray}
with $U(x_\mu ) \in SU(3), ~ F_\pi \cong 186Mev$, and $e$ is a
dimensionless constant.  The anomaly\cite{2} cannot be written as an
integral over space-time, but appears in the form:
\begin{eqnarray}
\Gamma = \frac{1}{240\pi^2} \int d \Sigma^{ijklm} Tr
(V_iV_jV_kV_lV_m)
\end{eqnarray}
with $V_j \equiv -U^{-1} \partial_j U, U \in SU(3)$, and $d
\Sigma^{ijklm}$ a volume element in an extended five-dimensional
space; the boundary of the integration region is compactified space-time.

Topological considerations enter as follows: for a given time $t$, the
matrix $U(t,x)$ is a mapping from $\mbox{\boldmath $R$}^3$ into
$SU(3)$.  The proper boundary conditions add the point at infinity to
three-space compactifying it to $S^3$.  As is familiar from Witten's
work,\cite{2} the equivalence classes of all such maps are classified by
the homotopy group $\pi_3 (SU(3)) = Z$, (the integral baryon number B).

For three-flavors the lowest energy B=1 soliton of the S'tW model is the
$3 \times 3$ matrix:
\begin{eqnarray}
\Sigma (r) = \left( \begin{array}{cc} \exp( \frac{2i}{F_\pi} \tau \cdot
\hat{r} F (r)) & 0 \\ 0 & 1 \end{array} \right),
\end{eqnarray}
where $\tau$ denotes the $2 \times 2$ isospin matrices and $F(r)$ is
determined from the
Euler-Lagrange equations.  The soliton $\Sigma$ has the symmetry
$SU(3)_{flavor} \times
SU(2)_{spin}$, where $SU(3)_{flavor}$ is realized by the adjoint action and
$SU(2)_{spin}$ by
transformations on {\bf r} (generators {\bf J}.  Eq.(4) shows a special
``{\bf K}-symmetry'' in
that $\Sigma$ is invariant under combined isospin-spin rotations:  ({\bf I}
+ {\bf
J})($\Sigma) = 0,$ and, moreover, the $SU(3)_{flavor}$ hypercharge
generator $Y$ also
leaves $\Sigma$ invariant: $[Y, \Sigma] = 0$.

The quantal eigenstates of the S'tW model are projected from the soliton
$\Sigma$, and are
{\it monopolar harmonics}\cite{7}, sections of a fiber bundle over the
coset manifold
$SU3/U1$.  These monopolar harmonics are specializations of the $SU3$
rotation matrices
over eight angles $(\alpha_1, i = 1 \ldots 8)$, with the angle $a_8$
determined by the
sectional map.  The $SU(3) \times SU(3)$ symmetry of the $SU(3)$ rotation
matrices
(generated by left and right actions on the $SU(3)$ manifold) reduces to
the symmetry
$SU(3)_{flavor} \times SU(2)_{spin}$ for the monopolar harmonics, with the
anomalous right
action hypercharge $Y_R \rightarrow N_c B/3$.

Can one construct a large-$N_c$ quark model unitarily equivalent to this
S'tW three- flavor
model?  Manohar\cite{4} was the first to attempt this and his model was
based on the
hedgehog quark:
\begin{eqnarray}
\left| \Sigma_q \right> \equiv  \left( \left| u \left> \otimes \right|
\downarrow \right> - \left| d \left> \otimes \right| \uparrow \right> \right)
\end{eqnarray}
where (as in the two-flavor case) $\mbox{\boldmath $K$} \left| \Sigma_q
\right> = 0,$
where $\mbox{\boldmath $K$} = \mbox{\boldmath $I$} + \mbox{\boldmath $J$}$.
 What
distinguishes this ket vector, $\left| \Sigma_q \right>$, from the
two-flavor case is the set
of allowed operations:  for the three-flavor case, one allows arbitrary
$SU(3)_{flavor}$ and
$SU(2)_{spin}$ transformations to act on $\left| \Sigma_q \right>$.
Transforming $\left|
\Sigma_q \right>$ by $R(g)$, where $g \in SU(3)_{flavor}$, carries $\left|
\Sigma_q
\right>$ to sufficiently many independent states to enable one to determine
the six basis
states of the defining irrep of $SU(3)_{flavor} \times SU(2)_{spin}.$

Consider now the anti-symmetrized color singlet $N_c$-hedgehog quark state:
\begin{eqnarray}
\left| \Sigma_q \right>_{N_c} \equiv \underbrace{ \left( \left| \Sigma_q
\left> \otimes
\left| \Sigma_q \right> \ldots \otimes \right| \Sigma_q \right>
\right)}_{N_c times}
\chi_ { color ~ singlet},
\end{eqnarray}
If we perform the transformation: $ \left| \Sigma_q \right> \rightarrow
R(g) \left|
\Sigma_q \right>$, $g \in SU(3)_{flavor}$ this induces a general
transformation on $\left|
\Sigma_q \right>_{N_c}$.  The states spanned by $\left\{ R(g) \left| \Sigma_q
\right>_{N_c} , g \in SU(3) \right\}$ are precisely the states of the
totally symmetric
irrep $[ N_c \dot{0} ]$ in $SU(6)$.  The $SU(3) \times SU(2)$ structure of
the irrep $[ N_c
\dot{0} ]$ in $SU(6)$ is well known\cite{8}:  Every irrep $[m_{13} m_{23} 0 ]
 \times [
m_{13} m_{23} ]$ in $SU(3) \times SU(2)$ occurs once and only once, subject
to lexicality
and the constraint that $m_{13} + m_{23} = N_c.$

To compare the baryonic states in this large $N_c$ quark model with the
baryonic states of
the three-flavor S'tW model we see that we must choose $N_c = 3k$, with $k$
an integer.
This comparison shows that there are serious discrepancies.  In fact {\it
only for} $N_c =3 $
{\it are the states the same}.  One finds that: (a) the spins in Manohar's
model for the baryon
tower are {\it not all half-integral}.  (b) the {\it multiplicity} of the
$SU(3)_{flavor}$
multiplets in the two systems is not the same, and (c) unlike the
two-flavor hedgehog
quark model, the multiplets for a given $N_c(=3k)$ do not contain the
multiplets for lower
$N_c(=3(k-1))$.  {\it This means that the physically important lowest
multiplets in the tower
will be obtained only with $N_c =3$, which prevents using the large-$N_c$
limit}.

To resolve these discrepancies, we remark that the proper three-flavor
hedgehog quark is not
defined by (6) but by the {\it local} [100] monopolar harmonic, which does
not exist globally, since such an object is {\it forbidden topologically}.
(More precisely these objects exist locally as confined triples).  Let us
assume
that this [100] state does exist locally; what would it look like?  Written as
a
matrix it would appear trivial, simply the $3 \times 3$ unit matrix.  But
recall that $SU(3)$ rotational wave functions realize the symmetry
$SU(3)_{left} \times SU(3)_{right}$ in which (conventionally) the left
generators {\it obey time-reversed commutation rules}.\cite{9}  Re-writing
the $3 \times 3$ unit matrix to accord with this basis, we find that
\begin{eqnarray} \left| \Sigma_0 \right> = \left(  \left| u \left> \otimes
\left| \downarrow \left> - \left| d \left> \otimes \right| \uparrow \right>
+ \right| s \right> \otimes \right| \rightarrow \right> \right).
\end{eqnarray} where $\left| \rightarrow \right>$ denotes a new
``sidewise", or spin 0, component.  Clearly the state $\left| \Sigma_0
\right>$ is invariant under an {\it octet} of $SU(3) \mbox{\boldmath
$K$}$-symmetry generators.

	We must justify this radical step impled by (7), but before we do so let
us remark that {\it
all of the difficulties noted above are resolved by this change.}

At first glance the introduction of a scalar $(S = 0)$ ``quark" is absurd;
there is no
experimental evidence for such an object.  But before we dismiss this idea
(which was, after all,
abstracted from the S'tW model) let us be more careful and examine the
right hypercharge (the
baryon number in the model).  We find $B= -2/3$.  {\it Thus the new} $S=0,
Y_R = B = -
\frac{2}{3}$ {\it state is an anti di-quark.}  There is credible evidence
that the di-quark\cite{6}
exists.\cite{10}

We conclude that the adjunction of the anti-diquark to define $\left|
\Sigma_0 \right>$ is
not completely unreasonable.

The resulting large-$N_c$ quark model is now straight-forward.  We must use
$N_c = 3k$ to
agree with the states of the three-flavor S'tW model.  The single ``quark"
basis implied by
$\left| \Sigma_0 \right>$ consists of the nine states in $SU(3)_{flavor} \times
SU(3)_{right}$.  Thus a baryonic structure composed of $N_c$ such quarks
consists of dim $[
N_c \dot{0} ]_{U9}$ symmetric states in $SU(9)$, since the color state is
an anti-symmetric
singlet in $SU(N_c)$.  It is important to realize that this $N_c$-quark
system consists of
``baryons" {\it with various baryonic charges} (right hypercharge $Y_R$ )
since the original
``quark" basis itself had two distinct baryonic charges $( \frac{1}{3}$ and
$-\frac{2}{3}$).
{\it Thus we must project from this system of} $N_c$ ``quarks" {\it the
states of} $Y_R = B
= 1$ {\it to accord with the $B = 1$ states of the three-flavor S'tW
model}.  It is easily
demonstrated that this projected $SU(3)_{flavor} \times SU(3)_{right}
\times SU(N_c)$
``quark" model for $N_c \rightarrow \infty$ is unitarily equivalent to the
three-flavor
S'tW model since the spectrum of states is {\it identical}.  Accordingly we
 propose to take
this new ``quark" model more seriously and examine its implications.

The results we have presented so far agree nicely with the S'tW model, and
are, group
theoretically, unassailable.  Despite this, a more critical analysis shows
there is a problem:
{\it following convention, we have used the symbol $N_c$ ambiguously, in
two distinct,
conflicting, ways}.  If the anomaly were absent, there would be no
ambiguity:  one simply
takes $SU(3)_{flavor} \times SU(2)_{spin} \times SU(N_c) ~ QCD$ in the 't
Hooft limit $N_c
\rightarrow \infty$ obtaining $SU(3)_{flavor} \times SU(2)_{spin}$
symmetry, defined ({\it
a la} Witten) on the manifold $SU(3)/U1$.  For clarity, let us henceforth
call 't Hooft's
parameter $N_{'t}$.  Now consider the anomaly.  First of all the anomaly
{\it cannot even be
defined} until we have constructed the limit $N_{'t} \rightarrow \infty$,
obtaining a smooth
manifold (over $N_{'t}$).  Thus the anomaly, which is quantized, introduces
a {\it new integral
parameter}, confusingly called $N_c$ again.  To avoid confusion let us call
this Witten's
parameter\cite{11} denoted by $N_W$.

The limit $N_{'t} \rightarrow \infty$ clearly eliminates the parameter
$N_{'t}$ from the
S'tW model, which means {\it that in the S'tW model color has been totally
confined and
disappears from the model.}  The anomaly makes a profound difference.
Witten's parameter
$N_W$ must be 3, which implies that right hypercharge measures the baryon
quantum
number B.  On purely topological grounds, {\it the anomaly forbids the
existence of S'tW
states with fractional B (that is, non-zero triality\cite{12}).}  {\it
Triples} of objects, with
composite triality zero, are, however, {\it not} forbidden {\it if confined
within the volume of the
Skyrmion} (or to even smaller distances).  Thus even though color has
disappeared we still
have a remnant:  quark confinement in triples.  (This beautifully
illustrates Gell-Mann's
intuition that quarks may be fictitious (unobservable) in his classic
pre-color paper.\cite{6})

For consistency with both the large-$N_{'t}$ quark model and the admissible
quark triples,
one must take the $N_{'t} \rightarrow \infty$ limit to run over the
integers $N_{'t} \equiv
0 ~ mod ~ 3$.  Thus we see the true relationship\cite{13} between 't
Hooft's parameter
$N_{'t}$ in the large-$N_{'t}$ quark model and Witten's anomaly parameter
$N_W$ is that
$N_{'t} \equiv t$ mod $N_W$ with $N_W = 3$.  (Here $t = 0, \pm 1$ is the
triality of the
$SU(3)_{flavor}$ irrep.)

Accordingly we abstract from the S'tW model the following structure:  a
quark soliton is a
{\it confined} `solution' to the S'tW model having $B = \frac{1}{3}$,
belonging to the {\it
broken} spectrum generating symmetry $U(9) \subset SU(3)_{flavor} \times
SU(3)_{right}
\subset SU(3)_{flavor} \times SU(2)_{spin} \times U1_{Y_R}$.  Projected
onto quantal
states, this $B = \frac{1}{3}$ system has an infinite tower of states:
$[100] \frac{1}{2}$,$ [220]
\frac{1}{2}$,$ [310] \frac{1}{2}$,$ [310] \frac{3}{2}, [400]
\frac{3}{2};\ldots$.  A finite subset of
these soliton quark states can be generated by $N_{'t} \equiv 1$ mod $3$
confined colorless
totally symmetrized light (mass $\sim N_{'t}^{-1}$) ``quarks", comprising
the states:
$[100] \frac{1}{2} \ldots,[N_{'t}00] ( \frac{2N_{'t}+1}{6} )$.  The
smallest such subset
$N_{'t} =3)$ are the {\it six} $B = 1/3$ pre-color Gell-Mann quarks: $[100]
1/2$, and, in
addition, the {\it three} (structurally essential) $B = -\frac{2}{3}$
``quarks" $[100] 0$, that
is, nine $[100]_{flavor} \times [100]_{right}$ states in all.  Composite
projected states of
three quark solitons are baryons (B =1).

Quark soliton composites are automatically equipped with a certain form of
interaction:  it is
easily shown\cite{14} that in the limit $N_{'t} = \infty$ {\it all
interactions are local in the
group (coset) manifold coordinates.}  This local interaction predicts
three-flavor
meson-baryon scattering\cite{15} for $N_{'t} = \infty$, generalizing the
earlier (two-flavor)
results of Mattis-Braaten.\cite{5}

The manifold coordinates (seven dimensional) are dual to the
representation-label space
(seven dimensional) so that finite $N_{'t}$ is to be associated with
non-locality which can be
modelled by group-theoretically defined tensor operators.  Expressed
differently the loss of
color degrees of freedom by explicit confinement is compensated by the
freedom to model,
group-theoretically, interactions in composite (large-$N_{'t})$ states.
The model we
propose implies structurally many features previously introduced
heuristically.  Among these
are: suppression of orbital angular momentum (quark-diquark baryonic
excited states), explicit
(`bag model') quark confinement, massless quarks, zero flavor triality, the
`spin-baryonic
charge' rule $(2S \equiv B ~ mod ~ 2)$ and suppression of nuclear
`hidden-color' effects.

The $SU(3)_{color}$ symmetry of QCD is a deep and profound organizing
principle in particle
physics.  It will be interesting to see if soliton quarks, which define a
{\it color-free} approach
to long-range, low energy QCD structures, can successfully approximate the real
meson-nuclear world.

\bigskip

\noindent {\bf Acknowledgements:}

We would like to thank Dr. Michael Mattis (LANL) for interesting us in
large-$N_c$ models
and for many clarifying discussions.  One of us (LPH) wishes to thank S.L.
Adler for his
hospitality at the IAS and the Monel Foundation for partial support.  This
work was
supported in part by grant DE-FG03-93ER40757.

\bigskip

\noindent $^{(a)}$ Permanent address:  School of Physics, Raymond and
Beverly Sackler
Faculty of Exact Sciences, Tel Aviv University, Ramat Aviv, Israel.


\begin{thebibliography}{99}
\bibitem{11} Witten had good reason to call this new parameter ``the number
of colors''.  He
showed that the anomaly implied an electromagnetic interaction which, to
agree with
experiment, forced the anomaly parameter $N_W$ to be 3, the physical number
of colors.
But we insist that Witten's anomaly parameter, $N_W$, is logically distinct
from
the parameter $(N_{'t} = \infty)$ used in defining the $SU(3)/U1$ manifold
itself.
\end{thebibliography}
\end{document}